\documentclass[twocolumn,12pt]{iopart}

\usepackage{graphicx}

\begin{document}

\title[]{Complete analysis of the maximally hyperentangled state via the weak cross-Kerr nonlinearity}
\author{Zhi Zeng$^{1,2,^*}$}
\address{{$^1$Institute of Signal Processing and Transmission, Nanjing University of Posts and Telecommunications, Nanjing, 210003, China}  \\
{$^2$Key Lab of Broadband Wireless Communication and Sensor Network Technology, Ministry of Education, Nanjing University of Posts and Telecommunications, Nanjing, 210003, China}}
\eads{\mailto{zengzhiphy@yeah.net}}
\vspace{10pt}

\begin{abstract}
We present a simple method for the complete analysis of maximally hyperentangled state in polarization and spatial-mode degrees of freedom assisted by the weak cross-Kerr nonlinearity. Our method not only can be used for two-photon hyperentangled Bell state analysis and three-photon hyperentangled Greenberger-Horne-Zeilinger (GHZ) state analysis, but also is suitable for $N$-photon hyperentangled GHZ state analysis. In our protocols, the bit information of hyperentanglement is read out via the nonlinear interaction, and the phase information is obtained by using linear optical element and single photon detector. This approach is achievable with the current technology, and will be useful for the practical high-capacity quantum communication schemes.
\end{abstract}

\section{Introduction}
Quantum entanglement is the key resource in quantum computation and quantum communication, and it has been widely used for quantum key distribution \cite{qkd1,qkd2}, quantum teleportation \cite{te}, quantum dense coding \cite{dense}, quantum entanglement swapping \cite{swap}, quantum secret sharing \cite{qss1,qss2,qss3} and quantum secure direct communication \cite{qsdc1,qsdc2,qsdc3,qsdc4}. Quantum simulation is very useful to the quantum science and technology. Recently, the efficient simulation of photosynthetic light harvesting has attracted much attention \cite{QS1,QS2,QS3}. In many quantum information processing tasks, the complete analysis of a set of orthogonal entangled states is an indispensable step to read out information \cite{AP1,AP2,AP3,AP4,AP5}. Among the all types of entangled state analysis, the Bell state analysis (BSA) for two-qubit system and Greenberger-Horne-Zeilinger (GHZ) state analysis (GSA) for multi-qubit system are mostly used. Research has showed that the linear-optics-based BSA and GSA are impossible, and the optimal success probability of them are 50\% and 25\%, respectively \cite{no1,no2}. An efficient way to deal with this problem is using hyperentanglement \cite{hyper1}, which is defined as entanglement in more than one degree of freedom (DOF). In 1998, Kwiat \emph{et al.} showed that the four polarization Bell states can be reliably identified just using linear optical elements, resorting to an additional DOF (time-bin or spatial-mode) \cite{momentum1}. In 2003, Walborn \emph{et al.} presented a simplified method for BSA assisted by auxiliary entanglement \cite{momentum2}. Subsequently, several other DOFs such as linear momentum DOF \cite{momentum3}, time-bin DOF \cite{time1,time2} and orbital angular momentum DOF \cite{oam} were used to accomplish polarization BSA in experiment. In 2013, Song \emph{et al.} proposed a scheme for $N$-photon GSA using hyperentanglement \cite{GSA1}. In 2014, frequency entanglement was utilized for the complete polarization BSA and GSA \cite{GSA2}. In addition to BSA and GSA, hyperentanglement also has many other important applications in quantum information processing, such as deterministic entanglement purification protocols \cite{hyper2,hyper3,hyper4,hyper5,hyper6}, hyperentanglement purification and concentration \cite{hyper7,hyper8,hyper9,hyper10,hyper11,hyper12}, hyper-parallel optical quantum computation \cite{hyper13,hyper14,hyper15,hyper16}, and high-efficiency quantum communication protocols \cite{QC1,QC2,QC3,QC4,QC5}.

Hyperentangled Bell state analysis (HBSA) and hyperentangled GHZ state analysis (HGSA) are essential to many high-capacity quantum communication schemes. However, it has been shown that the 16 hyperentangled Bell states in two DOFs only can be classified into 7 groups via linear optics, and thus the complete HBSA is not achieved \cite{no3,no4}. If the quantum nonlinear interaction is introduced, complete HBSA and HGSA can be both realized \cite{HBSA1,HGSA1,HBSA2,Htime,Hthree,HBSA3,HGSA2,HBSA4,HBSA5,HBSA6,HBSA7}. In 2010, Sheng \emph{et al.} proposed the first complete HBSA scheme for polarization and spatial-mode hyperentanglement based on the cross-Kerr nonlinearity \cite{HBSA1}. In 2012, Xia \emph{et al.} presented an efficient HGSA scheme using a similar principle \cite{HGSA1}. In 2016, Li \emph{et al.} proposed the self-assisted complete hyperentangled state analysis scheme for HBSA and HGSA, in which the discrimination process can be greatly simplified \cite{HBSA2}. In the past decade, the quantum-dot spin in optical microcavity and nitrogen-vacancy center in resonator have also been used for hyperentangled state analysis \cite{HBSA3,HGSA2,HBSA4,HBSA5,HBSA6,HBSA7}. An alternative method for complete HBSA and HGSA is using the auxiliary entanglement in other DOFs of photon \cite{HBSA8,HBSA9,HBSA10}. In 2017, Li \emph{et al.} showed that the 16 hyperentangled Bell states can be separated into 12 groups with the help of an auxiliary known entangled state in the third DOF \cite{HBSA8}. In 2019, Wang \emph{et al.} proposed an efficient HBSA scheme for polarization and the first longitudinal momentum hyperentanglement, by using a fixed hyperentangled Bell state in frequency and the second longitudinal momentum DOFs \cite{HBSA9}.

In this paper, we present an efficient and simple method for the complete analysis of hyperentangled Bell state and hyperentangled GHZ state in polarization and spatial-mode DOFs via the weak cross-Kerr nonlinearity. The bit information of hyperentangled state are obtained through the measurement on coherent probe state, and the relative phase information are determined by using linear optical element and single photon detector. The principle of our schemes is quite different from the self-assisted mechanism proposed by Li \emph{et al.} in 2016 \cite{HBSA2}, in which the cross-Kerr nonlinearity is also utilized. Our method can be directly extended to the complete $N$-photon HGSA, and will have useful applications in high-capacity long-distance quantum communication.

\section{Complete analysis of hyperentangled Bell state}
In this section, we first introduce the photon number quantum nondemolition detector (QND), which is constructed by the weak cross-Kerr nonlinearity. Then, we will detailly demonstrate our complete HBSA scheme for the hyperentangled photons in polarization and spatial-mode DOFs.

\subsection{The principle of photon number QND}
The cross-Kerr effect is an interaction between a signal state $|\psi\rangle_s$ and a probe coherent state $|\alpha\rangle_p$ in a nonlinear medium with the Hamiltonian \cite{Kerr1}
\begin{eqnarray}
H = \hbar\chi a^\dag_sa_sa^\dag_pa_p.
\end{eqnarray}
$a^\dag_s$ ($a^\dag_p$) and $a_s$ ($a_p$) are the creation and destruction operators of the signal (probe) state, respectively. $\chi$ is the coupling strength of the nonlinearity, which depends on the property of nonlinear material. Supposing the signal state is a superposition of the Fock state as
\begin{eqnarray}
|\psi\rangle_s = a|0\rangle_s + b|1\rangle_s + c|2\rangle_s,
\end{eqnarray}
the effect of the cross-Kerr nonlinearity can be described as
\begin{eqnarray}
U|\psi\rangle_s|\alpha\rangle_p = a|0\rangle_s|\alpha\rangle_p + b|1\rangle_s|\alpha e^{i\theta}\rangle_p + c|2\rangle_s|\alpha e^{2i\theta}\rangle_p.
\end{eqnarray}
Here, $\theta=\chi t$ and $t$ is the interaction time. The coherent probe state will pick up a phase shift, which is proportional to the number of photon in the signal state. Using the $X$ quadrature measurement, the information of phase shift can be obtained, and then the number of photon in the signal state can be measured without destroying the photons.

\subsection{Complete HBSA for polarization-spatial-mode hyperentanglement}
The two-photon hyperentangled Bell state in polarization and spatial-mode DOFs can be written as
\begin{eqnarray}
|\Psi\rangle_{AB} = |\zeta\rangle_{P} \otimes |\eta\rangle_{S}.
\end{eqnarray}
Here $A$ and $B$ represent the two entangled photons, and the subscripts $P$ and $S$ denote the polarization and spatial-mode DOFs, respectively. $|\zeta\rangle_{P}$ can be one of the four polarization Bell states,
\begin{eqnarray}
|\phi^{\pm}\rangle_{P} = \frac{1}{\sqrt 2} (|HH\rangle \pm |VV\rangle)_{AB}, \nonumber \\
|\psi^{\pm}\rangle_{P} = \frac{1}{\sqrt 2} (|HV\rangle \pm |VH\rangle)_{AB}.
\end{eqnarray}
$H$ and $V$ are the horizontal and vertical polarization states, respectively. $|\eta\rangle_{S}$ can be one of the four spatial-mode Bell states,
\begin{eqnarray}
|\phi^{\pm}\rangle_{S} = \frac{1}{\sqrt 2} (|a_{1}b_{1}\rangle \pm |a_{2}b_{2}\rangle)_{AB}, \nonumber \\
|\psi^{\pm}\rangle_{S} = \frac{1}{\sqrt 2} (|a_{1}b_{2}\rangle \pm |a_{2}b_{1}\rangle)_{AB}.
\end{eqnarray}
Here $a_{1}$ ($b_{1}$) and $a_{2}$ ($b_{2}$) are the two different possible spatial-modes of photon $A$ ($B$). The 16 hyperentangled states contain two kinds of information, namely the bit information and the relative phase information. In our proposal, the bit information is identified through the two QNDs, which are accessible with the current technology. The relative phase information is read out assisted by the linear optical element and single photon detector. Thus, the 16 hyperentangeld Bell states in two DOFs can be completely distinguished.

\begin{figure}
\centering
\includegraphics*[width=0.8\textwidth]{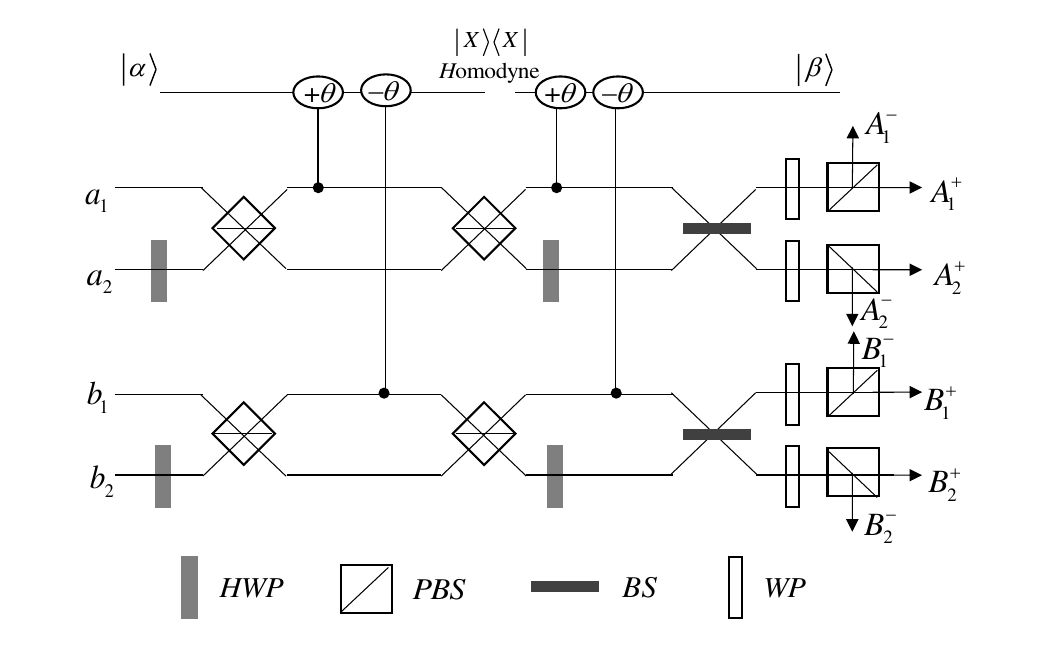}
\caption{Schematic diagram of our complete HBSA scheme for polarization and spatial-mode hyperentanglement. The half-wave plates (HWPs) implement the bit-flip operation [$|H\rangle\leftrightarrow|V\rangle$] on the polarization state of photon. The polarization beam splitters (PBSs) transmit the horizontal polarized photon with its path switched, and reflect the vertical one with the path unchanged. The cross-Kerr nonlinear interactions generate the phase shift of $\pm\theta$ on the coherent states $|\alpha\rangle$ and $|\beta\rangle$ if photons appear in the corresponding spatial-modes. The beam splitters (BSs) accomplish the Hadamard operation [$|a_1\rangle\rightarrow\frac{1}{\sqrt 2}(|a_1\rangle + |a_2\rangle), |a_2\rangle\rightarrow\frac{1}{\sqrt 2}(|a_1\rangle - |a_2\rangle), |b_1\rangle\rightarrow\frac{1}{\sqrt 2}(|b_1\rangle + |b_2\rangle), |b_2\rangle\rightarrow\frac{1}{\sqrt 2}(|b_1\rangle - |b_2\rangle)$] on the spatial-mode state of photon. The wave plates (WPs) perform the Hadamard operation [$|H\rangle\rightarrow\frac{1}{\sqrt 2}(|H\rangle + |V\rangle), |V\rangle\rightarrow\frac{1}{\sqrt 2}(|H\rangle - |V\rangle)$] on the polarization state of photon. Single photon detectors are required to realize the polarization measurement in $\{|H\rangle,|V\rangle\}$ basis. With this device, the 16 hyperentangled Bell states can be completely distinguished.}
\end{figure}

The setup of our HBSA scheme for polarization-spatial-mode hyperentanglement is shown in Fig. 1. After the two photons pass through HWPs and PBSs, interact with the coherent state $|\alpha\rangle$, pass through the PBSs and HWPs again, then the quantum state of the collective system composed of hyperentanglement and coherent state evolves as
\begin{eqnarray}
|\phi^{\pm}\rangle_{P}|\eta\rangle_{S}|\alpha\rangle &\rightarrow& |\phi^{\pm}\rangle_{P}|\eta\rangle_{S}|\alpha\rangle, \nonumber \\
|\psi^{\pm}\rangle_{P}|\eta\rangle_{S}|\alpha\rangle &\rightarrow& |\psi^{\pm}\rangle_{P}|\eta\rangle_{S}|\alpha e^{\pm i\theta}\rangle.
\end{eqnarray}
We can find that the entanglement in spatial-mode DOF is invariant during the evolution. It should be noted that the above derivation follows from the fact that the $X$-quadrature measurement on coherent state is set up only to distinguish the phase 0 from $\pm\theta$. Thus, the bit information of polarization entangled state can be obtained after the $X$-quadrature measurement on $|\alpha\rangle$. Subsequently, the photons $A$ and $B$ will interact with coherent state $|\beta\rangle$, and the evolution of the collective system is
\begin{eqnarray}
|\zeta\rangle_{P}|\phi^{\pm}\rangle_{S}|\beta\rangle &\rightarrow& |\zeta\rangle_{P}|\phi^{\pm}\rangle_{S}|\beta\rangle, \nonumber \\
|\zeta\rangle_{P}|\psi^{\pm}\rangle_{S}|\beta\rangle &\rightarrow& |\zeta\rangle_{P}|\psi^{\pm}\rangle_{S}|\beta e^{\pm i\theta}\rangle.
\end{eqnarray}
It is shown that the polarization entanglement is invariant in the interaction, and we can determine the bit information of spatial-mode entanglement by utilizing the $X$-quadrature measurement on $|\beta\rangle$. Therefore, the 16 hyperentangled Bell states can be classified into four groups with the help of the coherent states $|\alpha\rangle$ and $|\beta\rangle$, as shown in Table 1.

\begin{table}
\centering\caption{Corresponding relations between the initial states and phase shifts of the two coherent states.}
\begin{tabular}{cc ccccccccccc}
\hline
Initial states & & & & $|\alpha\rangle$ & & & & $|\beta\rangle$ \\
\hline
$|\phi^{+}\rangle_P|\phi^{\pm}\rangle_S$, $|\phi^{-}\rangle_P|\phi^{\pm}\rangle_S$. &&&& $0$ &&&& $0$ \\
$|\phi^{+}\rangle_P|\psi^{\pm}\rangle_S$, $|\phi^{-}\rangle_P|\psi^{\pm}\rangle_S$. &&&& $0$ &&&& $\pm\theta$ \\
$|\psi^{+}\rangle_P|\phi^{\pm}\rangle_S$, $|\psi^{-}\rangle_P|\phi^{\pm}\rangle_S$. &&&& $\pm\theta$ &&&& $0$ \\
$|\psi^{+}\rangle_P|\psi^{\pm}\rangle_S$, $|\psi^{-}\rangle_P|\psi^{\pm}\rangle_S$. &&&& $\pm\theta$ &&&& $\pm\theta$ \\
\hline
\end{tabular}
\end{table}

In each group, there are four states that have the same bit information but different relative phase information, and these four states can be distinguished by using the BSs and WPs. Here, we take the second group in Table 1 as an example to illustrate. After passing through BSs and WPs, the evolution of the hyperentangled state is
\begin{eqnarray}
|\phi^{+}\rangle_{P}|\psi^{+}\rangle_{S} \rightarrow \frac{1}{2} (|HH\rangle + |VV\rangle) \otimes (|a_{1}b_{1}\rangle - |a_{2}b_{2}\rangle)_{AB}, \nonumber \\
|\phi^{+}\rangle_{P}|\psi^{-}\rangle_{S} \rightarrow \frac{1}{2} (|HH\rangle + |VV\rangle) \otimes (|a_{1}b_{2}\rangle - |a_{2}b_{1}\rangle)_{AB}, \nonumber \\
|\phi^{-}\rangle_{P}|\psi^{+}\rangle_{S} \rightarrow \frac{1}{2} (|HV\rangle + |VH\rangle) \otimes (|a_{1}b_{1}\rangle - |a_{2}b_{2}\rangle)_{AB}, \nonumber \\
|\phi^{-}\rangle_{P}|\psi^{-}\rangle_{S} \rightarrow \frac{1}{2} (|HV\rangle + |VH\rangle) \otimes (|a_{1}b_{2}\rangle - |a_{2}b_{1}\rangle)_{AB}.
\end{eqnarray}
According to the detection responses of single photon detectors, the above four states can be determinately identified. In Table 2, we list the possible initial hyperentangled Bell states with their corresponding detection results. Using Table 1 and Table 2 simultaneously, we can unambiguously distinguish the 16 hyperentangled Bell states in polarization and spatial-mode DOFs.

\begin{table}
\centering\caption{Corresponding relations between the initial 16 hyperentangled Bell states and the possible detection results of single photon detectors.}
\begin{tabular}{cc ccccccccccc}
\hline
Initial states & & & & Possible detections \\
\hline
$|\phi^+\rangle_P|\phi^+\rangle_S$,$|\phi^+\rangle_P|\psi^+\rangle_S$,$|\psi^+\rangle_P|\phi^+\rangle_S$,$|\psi^+\rangle_P|\psi^+\rangle_S$. &&&& $A^{+}_1B^{+}_1$, $A^{-}_1B^{-}_1$, $A^{+}_2B^{+}_2$, $A^{-}_2B^{-}_2$.   \\
$|\phi^+\rangle_P|\phi^-\rangle_S$,$|\phi^+\rangle_P|\psi^-\rangle_S$,$|\psi^+\rangle_P|\phi^-\rangle_S$,$|\psi^+\rangle_P|\psi^-\rangle_S$. &&&& $A^{+}_1B^{+}_2$, $A^{-}_1B^{-}_2$, $A^{+}_2B^{+}_1$, $A^{-}_2B^{-}_1$.   \\
$|\phi^-\rangle_P|\phi^+\rangle_S$,$|\phi^-\rangle_P|\psi^+\rangle_S$,$|\psi^-\rangle_P|\phi^+\rangle_S$,$|\psi^-\rangle_P|\psi^+\rangle_S$. &&&& $A^{+}_1B^{-}_1$, $A^{-}_1B^{+}_1$, $A^{+}_2B^{-}_2$, $A^{-}_2B^{+}_2$.   \\
$|\phi^-\rangle_P|\phi^-\rangle_S$,$|\phi^-\rangle_P|\psi^-\rangle_S$,$|\psi^-\rangle_P|\phi^-\rangle_S$,$|\psi^-\rangle_P|\psi^-\rangle_S$. &&&& $A^{+}_1B^{-}_2$, $A^{-}_1B^{+}_2$, $A^{+}_2B^{-}_1$, $A^{-}_2B^{+}_1$.   \\

\hline
\end{tabular}
\end{table}

\section{Complete analysis of hyperentangled Greenberger-Horne-Zeilinger state}
The $N$-photon hyperentangled GHZ state in polarization and spatial-mode DOFs can be written as
\begin{eqnarray}
|\Phi\rangle_{AB\cdots N} = |\Phi\rangle_{P} \otimes |\Phi\rangle_{S},
\end{eqnarray}
in which $A,B\cdots N$ represent the $N$ entangled photons. $|\Phi\rangle_{P}$ ($|\Phi\rangle_{S}$) can be one of the $2^N$ polarization (spatial-mode) GHZ states,
\begin{eqnarray}
|\Phi^{\pm}_{ij\cdots k}\rangle_{P(S)} = \frac{1}{\sqrt 2} (|ij\cdots k\rangle \pm |\bar{i}\bar{j}\cdots \bar{k}\rangle)_{AB\cdots N}.
\end{eqnarray}
Here, $i,j\cdots k \in \{0,1\}$ and $m = 1-\bar{m} (m = i,j\cdots k)$. For the polarization DOF, $|0\rangle \equiv |H\rangle$ and $|1\rangle \equiv |V\rangle$. For the spatial-mode DOF, $|0\rangle \equiv |x_1\rangle$ and $|1\rangle \equiv |x_2\rangle (x=a,b\cdots n)$. $|x_1\rangle$ and $|x_2\rangle$ are the two possible spatial-modes of photon $X (X=A,B\cdots N)$. Taking into account polarization and spatial-mode DOFs together, there are $4^N$ hyperentangled GHZ states for the $N$-photon system, which will be distinguished in the following text. We first describe the complete HGSA scheme for three-photon system, and then generalize the scheme to $N$-photon HGSA directly.

\subsection{Complete HGSA for the three-photon hyperentanglement}
When $N = 3$, one of the 64 hyperentangled GHZ states in two DOFs can be written as
\begin{eqnarray}
|\Phi^{+}_{000}\rangle_{P} \otimes |\Phi^{+}_{000}\rangle_{S} = \frac{1}{\sqrt 2} (|HHH\rangle + |VVV\rangle) \otimes \frac{1}{\sqrt 2} (|a_1b_1c_1\rangle + |a_2b_2c_2\rangle)_{ABC}.  \nonumber \\
\end{eqnarray}

\begin{figure}
\centering
\includegraphics*[width=0.8\textwidth]{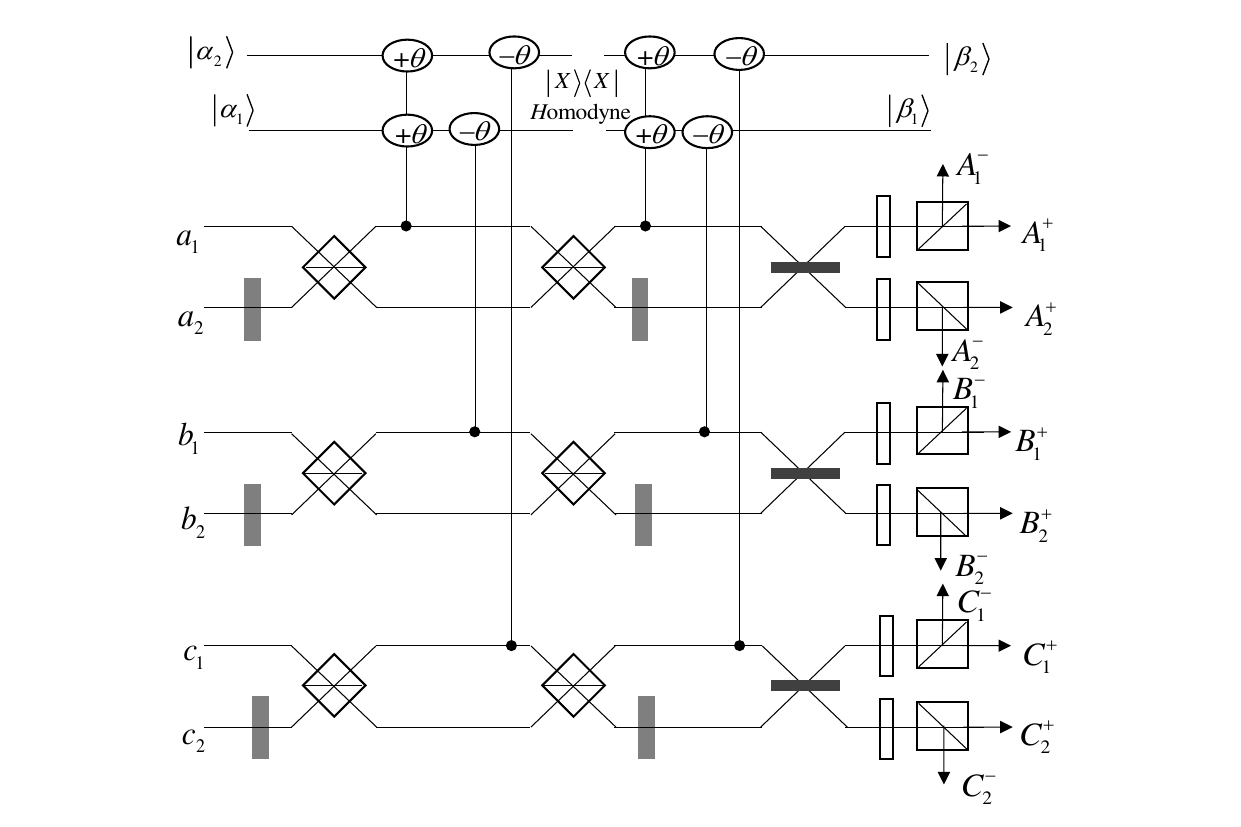}
\caption{Schematic diagram of our complete HGSA scheme for the three-photon hyperentanglement. Four coherent states are utilized to determine the bit information of hyperentangled GHZ states, then the relative phase information can be obtained by using linear optical elements and single photon detectors. With this device, the 64 hyperentangled GHZ states in polarization and spatial-mode DOFs can be completely distinguished.}
\end{figure}

In our three-photon HGSA scheme, four QNDs are required to get the bit information of hyperentanglement, as shown in Fig. 2. After photons $A$, $B$ and $C$ pass through HWPs and PBSs, interact with the coherent states $|\alpha_1\rangle$ and $|\alpha_2\rangle$, pass through the PBSs and HWPs again, the quantum state of the collective system evolves as
\begin{eqnarray}
|\Phi^{\pm}_{000}\rangle_{P}|\Phi\rangle_{S}|\alpha_1\rangle|\alpha_2\rangle &\rightarrow& |\Phi^{\pm}_{000}\rangle_{P}|\Phi\rangle_{S}|\alpha_1\rangle|\alpha_2\rangle,   \nonumber \\
|\Phi^{\pm}_{001}\rangle_{P}|\Phi\rangle_{S}|\alpha_1\rangle|\alpha_2\rangle &\rightarrow& |\Phi^{\pm}_{001}\rangle_{P}|\Phi\rangle_{S}|\alpha_1\rangle|\alpha_2 e^{\pm i\theta}\rangle,  \nonumber \\
|\Phi^{\pm}_{010}\rangle_{P}|\Phi\rangle_{S}|\alpha_1\rangle|\alpha_2\rangle &\rightarrow& |\Phi^{\pm}_{010}\rangle_{P}|\Phi\rangle_{S}|\alpha_1e^{\pm i\theta}\rangle|\alpha_2\rangle,  \nonumber \\
|\Phi^{\pm}_{100}\rangle_{P}|\Phi\rangle_{S}|\alpha_1\rangle|\alpha_2\rangle &\rightarrow& |\Phi^{\pm}_{100}\rangle_{P}|\Phi\rangle_{S}|\alpha_1e^{\pm i\theta}\rangle|\alpha_2e^{\pm i\theta}\rangle.
\end{eqnarray}
The spatial-mode entanglement is invariant during the evolution, and the bit information of polarization entanglement can be obtained assisted by the measurement on the two coherent states. After the three photons interact with the coherent states $|\beta_1\rangle$ and $|\beta_2\rangle$, the quantum state of the collective system evolves as
\begin{eqnarray}
|\Phi\rangle_{P}|\Phi^{\pm}_{000}\rangle_{S}|\beta_1\rangle|\beta_2\rangle &\rightarrow& |\Phi\rangle_{P}|\Phi^{\pm}_{000}\rangle_{S}|\beta_1\rangle|\beta_2\rangle,   \nonumber \\
|\Phi\rangle_{P}|\Phi^{\pm}_{001}\rangle_{S}|\beta_1\rangle|\beta_2\rangle &\rightarrow& |\Phi\rangle_{P}|\Phi^{\pm}_{001}\rangle_{S}|\beta_1\rangle|\beta_2 e^{\pm i\theta}\rangle,  \nonumber \\
|\Phi\rangle_{P}|\Phi^{\pm}_{010}\rangle_{S}|\beta_1\rangle|\beta_2\rangle &\rightarrow& |\Phi\rangle_{P}|\Phi^{\pm}_{010}\rangle_{S}|\beta_1e^{\pm i\theta}\rangle|\beta_2\rangle,  \nonumber \\
|\Phi\rangle_{P}|\Phi^{\pm}_{100}\rangle_{S}|\beta_1\rangle|\beta_2\rangle &\rightarrow& |\Phi\rangle_{P}|\Phi^{\pm}_{100}\rangle_{S}|\beta_1e^{\pm i\theta}\rangle|\beta_2e^{\pm i\theta}\rangle.
\end{eqnarray}
It is easy to find that the polarization entanglement is unchanged, and the bit information of spatial-mode entanglement is accessible. With the help of the four QNDs, the 64 hyperentangled GHZ states thus can be divided into 16 groups, as shown in Table 3.

\begin{table}
\centering\caption{Corresponding relations between the initial states and phase shifts of the four coherent states.}
\begin{tabular}{cc ccccc ccccc ccccc ccccc}
\hline
Initial states & & & & $|\alpha_1\rangle$ & & & & $|\alpha_2\rangle$ & & & & $|\beta_1\rangle$ & & & & $|\beta_2\rangle$ \\
\hline
$|\Phi^{+}_{000}\rangle_P|\Phi^{\pm}_{000}\rangle_S$, $|\Phi^{-}_{000}\rangle_P|\Phi^{\pm}_{000}\rangle_S$. &&&& $0$ &&&& $0$ &&&& $0$ &&&& $0$  \\
$|\Phi^{+}_{000}\rangle_P|\Phi^{\pm}_{001}\rangle_S$, $|\Phi^{-}_{000}\rangle_P|\Phi^{\pm}_{001}\rangle_S$. &&&& $0$ &&&& $0$ &&&& $0$ &&&& $\pm\theta$  \\
$|\Phi^{+}_{000}\rangle_P|\Phi^{\pm}_{010}\rangle_S$, $|\Phi^{-}_{000}\rangle_P|\Phi^{\pm}_{010}\rangle_S$. &&&& $0$ &&&& $0$ &&&& $\pm\theta$ &&&& $0$  \\
$|\Phi^{+}_{000}\rangle_P|\Phi^{\pm}_{100}\rangle_S$, $|\Phi^{-}_{000}\rangle_P|\Phi^{\pm}_{100}\rangle_S$. &&&& $0$ &&&& $0$ &&&& $\pm\theta$ &&&& $\pm\theta$  \\

$|\Phi^{+}_{001}\rangle_P|\Phi^{\pm}_{000}\rangle_S$, $|\Phi^{-}_{001}\rangle_P|\Phi^{\pm}_{000}\rangle_S$. &&&& $0$ &&&& $\pm\theta$ &&&& $0$ &&&& $0$  \\
$|\Phi^{+}_{001}\rangle_P|\Phi^{\pm}_{001}\rangle_S$, $|\Phi^{-}_{001}\rangle_P|\Phi^{\pm}_{001}\rangle_S$. &&&& $0$ &&&& $\pm\theta$ &&&& $0$ &&&& $\pm\theta$  \\
$|\Phi^{+}_{001}\rangle_P|\Phi^{\pm}_{010}\rangle_S$, $|\Phi^{-}_{001}\rangle_P|\Phi^{\pm}_{010}\rangle_S$. &&&& $0$ &&&& $\pm\theta$ &&&& $\pm\theta$ &&&& $0$  \\
$|\Phi^{+}_{001}\rangle_P|\Phi^{\pm}_{100}\rangle_S$, $|\Phi^{-}_{001}\rangle_P|\Phi^{\pm}_{100}\rangle_S$. &&&& $0$ &&&& $\pm\theta$ &&&& $\pm\theta$ &&&& $\pm\theta$  \\

$|\Phi^{+}_{010}\rangle_P|\Phi^{\pm}_{000}\rangle_S$, $|\Phi^{-}_{010}\rangle_P|\Phi^{\pm}_{000}\rangle_S$. &&&& $\pm\theta$ &&&& $0$ &&&& $0$ &&&& $0$  \\
$|\Phi^{+}_{010}\rangle_P|\Phi^{\pm}_{001}\rangle_S$, $|\Phi^{-}_{010}\rangle_P|\Phi^{\pm}_{001}\rangle_S$. &&&& $\pm\theta$ &&&& $0$ &&&& $0$ &&&& $\pm\theta$  \\
$|\Phi^{+}_{010}\rangle_P|\Phi^{\pm}_{010}\rangle_S$, $|\Phi^{-}_{010}\rangle_P|\Phi^{\pm}_{010}\rangle_S$. &&&& $\pm\theta$ &&&& $0$ &&&& $\pm\theta$ &&&& $0$  \\
$|\Phi^{+}_{010}\rangle_P|\Phi^{\pm}_{100}\rangle_S$, $|\Phi^{-}_{010}\rangle_P|\Phi^{\pm}_{100}\rangle_S$. &&&& $\pm\theta$ &&&& $0$ &&&& $\pm\theta$ &&&& $\pm\theta$  \\

$|\Phi^{+}_{100}\rangle_P|\Phi^{\pm}_{000}\rangle_S$, $|\Phi^{-}_{100}\rangle_P|\Phi^{\pm}_{000}\rangle_S$. &&&& $\pm\theta$ &&&& $\pm\theta$ &&&& $0$ &&&& $0$  \\
$|\Phi^{+}_{100}\rangle_P|\Phi^{\pm}_{001}\rangle_S$, $|\Phi^{-}_{100}\rangle_P|\Phi^{\pm}_{001}\rangle_S$. &&&& $\pm\theta$ &&&& $\pm\theta$ &&&& $0$ &&&& $\pm\theta$  \\
$|\Phi^{+}_{100}\rangle_P|\Phi^{\pm}_{010}\rangle_S$, $|\Phi^{-}_{100}\rangle_P|\Phi^{\pm}_{010}\rangle_S$. &&&& $\pm\theta$ &&&& $\pm\theta$ &&&& $\pm\theta$ &&&& $0$  \\
$|\Phi^{+}_{100}\rangle_P|\Phi^{\pm}_{100}\rangle_S$, $|\Phi^{-}_{100}\rangle_P|\Phi^{\pm}_{100}\rangle_S$. &&&& $\pm\theta$ &&&& $\pm\theta$ &&&& $\pm\theta$ &&&& $\pm\theta$  \\
\hline
\end{tabular}
\end{table}

In each group, there are four states with the same bit information but different relative phase information, which will be distinguished in the next step. We take the second group in Table 3 as an example to illustrate, and the evolution of the hyperentanglement after passing through BSs and WPs will be
\begin{eqnarray}
|\Phi^{+}_{000}\rangle_{P}|\Phi^{+}_{001}\rangle_{S} \rightarrow &&\frac{1}{4} (|HHH\rangle + |HVV\rangle + |VHV\rangle + |VVH\rangle)  \nonumber \\ &&\otimes (|a_{1}b_{1}c_{1}\rangle - |a_{1}b_{2}c_{2}\rangle - |a_{2}b_{1}c_{2}\rangle +|a_{2}b_{2}c_{1}\rangle)_{ABC}, \nonumber \\
|\Phi^{+}_{000}\rangle_{P}|\Phi^{-}_{001}\rangle_{S} \rightarrow &&\frac{1}{4} (|HHH\rangle + |HVV\rangle + |VHV\rangle + |VVH\rangle) \nonumber \\ &&\otimes (|a_{1}b_{1}c_{2}\rangle - |a_{1}b_{2}c_{1}\rangle - |a_{2}b_{1}c_{1}\rangle +|a_{2}b_{2}c_{2}\rangle)_{ABC},  \nonumber \\
|\Phi^{-}_{000}\rangle_{P}|\Phi^{+}_{001}\rangle_{S} \rightarrow &&\frac{1}{4} (|HHV\rangle + |HVH\rangle + |VHH\rangle + |VVV\rangle) \nonumber \\ &&\otimes (|a_{1}b_{1}c_{1}\rangle - |a_{1}b_{2}c_{2}\rangle - |a_{2}b_{1}c_{2}\rangle +|a_{2}b_{2}c_{1}\rangle)_{ABC},   \nonumber \\
|\Phi^{-}_{000}\rangle_{P}|\Phi^{-}_{001}\rangle_{S} \rightarrow &&\frac{1}{4} (|HHV\rangle + |HVH\rangle + |VHH\rangle + |VVV\rangle) \nonumber \\ &&\otimes (|a_{1}b_{1}c_{2}\rangle - |a_{1}b_{2}c_{1}\rangle - |a_{2}b_{1}c_{1}\rangle +|a_{2}b_{2}c_{2}\rangle)_{ABC}.
\end{eqnarray}
These four states are distinguishable through the detection results of single photon detectors, with which all the 64 hyperentangled GHZ states can be classified into four groups, as shown in Table 4. There are 16 hyperentangled states in each group, and we can perfectly identify them with the help of Table 3. Therefore, the complete HGSA scheme for three-photon system is achievable by using Table 3 and Table 4 simultaneously.

\begin{table}
\centering\caption{The 64 hyperentangled GHZ states can be classified into four groups based on the detection results of single photon detectors.}
\begin{tabular}{cc ccccccccccc}
\hline
Group &  Initial states \\
\hline
$1$ & $|\Phi^{+}_{000}\rangle_P|\Phi^{+}_{000}\rangle_S,|\Phi^{+}_{000}\rangle_P|\Phi^{+}_{001}\rangle_S,$
      $|\Phi^{+}_{000}\rangle_P|\Phi^{+}_{010}\rangle_S,|\Phi^{+}_{000}\rangle_P|\Phi^{+}_{100}\rangle_S,$ \\&
      $|\Phi^{+}_{001}\rangle_P|\Phi^{+}_{000}\rangle_S,|\Phi^{+}_{001}\rangle_P|\Phi^{+}_{001}\rangle_S,$
      $|\Phi^{+}_{001}\rangle_P|\Phi^{+}_{010}\rangle_S,|\Phi^{+}_{001}\rangle_P|\Phi^{+}_{100}\rangle_S,$ \\ &
      $|\Phi^{+}_{010}\rangle_P|\Phi^{+}_{000}\rangle_S,|\Phi^{+}_{010}\rangle_P|\Phi^{+}_{001}\rangle_S,$
      $|\Phi^{+}_{010}\rangle_P|\Phi^{+}_{010}\rangle_S,|\Phi^{+}_{010}\rangle_P|\Phi^{+}_{100}\rangle_S,$ \\ &
      $|\Phi^{+}_{100}\rangle_P|\Phi^{+}_{000}\rangle_S,|\Phi^{+}_{100}\rangle_P|\Phi^{+}_{001}\rangle_S,$
      $|\Phi^{+}_{100}\rangle_P|\Phi^{+}_{010}\rangle_S,|\Phi^{+}_{100}\rangle_P|\Phi^{+}_{100}\rangle_S.$ \\

$2$ & $|\Phi^{+}_{000}\rangle_P|\Phi^{-}_{000}\rangle_S,|\Phi^{+}_{000}\rangle_P|\Phi^{-}_{001}\rangle_S,$
      $|\Phi^{+}_{000}\rangle_P|\Phi^{-}_{010}\rangle_S,|\Phi^{+}_{000}\rangle_P|\Phi^{-}_{100}\rangle_S,$ \\&
      $|\Phi^{+}_{001}\rangle_P|\Phi^{-}_{000}\rangle_S,|\Phi^{+}_{001}\rangle_P|\Phi^{-}_{001}\rangle_S,$
      $|\Phi^{+}_{001}\rangle_P|\Phi^{-}_{010}\rangle_S,|\Phi^{+}_{001}\rangle_P|\Phi^{-}_{100}\rangle_S,$ \\ &
      $|\Phi^{+}_{010}\rangle_P|\Phi^{-}_{000}\rangle_S,|\Phi^{+}_{010}\rangle_P|\Phi^{-}_{001}\rangle_S,$
      $|\Phi^{+}_{010}\rangle_P|\Phi^{-}_{010}\rangle_S,|\Phi^{+}_{010}\rangle_P|\Phi^{-}_{100}\rangle_S,$ \\ &
      $|\Phi^{+}_{100}\rangle_P|\Phi^{-}_{000}\rangle_S,|\Phi^{+}_{100}\rangle_P|\Phi^{-}_{001}\rangle_S,$
      $|\Phi^{+}_{100}\rangle_P|\Phi^{-}_{010}\rangle_S,|\Phi^{+}_{100}\rangle_P|\Phi^{-}_{100}\rangle_S.$ \\

$3$ & $|\Phi^{-}_{000}\rangle_P|\Phi^{+}_{000}\rangle_S,|\Phi^{-}_{000}\rangle_P|\Phi^{+}_{001}\rangle_S,$
      $|\Phi^{-}_{000}\rangle_P|\Phi^{+}_{010}\rangle_S,|\Phi^{-}_{000}\rangle_P|\Phi^{+}_{100}\rangle_S,$ \\&
      $|\Phi^{-}_{001}\rangle_P|\Phi^{+}_{000}\rangle_S,|\Phi^{-}_{001}\rangle_P|\Phi^{+}_{001}\rangle_S,$
      $|\Phi^{-}_{001}\rangle_P|\Phi^{+}_{010}\rangle_S,|\Phi^{-}_{001}\rangle_P|\Phi^{+}_{100}\rangle_S,$ \\ &
      $|\Phi^{-}_{010}\rangle_P|\Phi^{+}_{000}\rangle_S,|\Phi^{-}_{010}\rangle_P|\Phi^{+}_{001}\rangle_S,$
      $|\Phi^{-}_{010}\rangle_P|\Phi^{+}_{010}\rangle_S,|\Phi^{-}_{010}\rangle_P|\Phi^{+}_{100}\rangle_S,$ \\ &
      $|\Phi^{-}_{100}\rangle_P|\Phi^{+}_{000}\rangle_S,|\Phi^{-}_{100}\rangle_P|\Phi^{+}_{001}\rangle_S,$
      $|\Phi^{-}_{100}\rangle_P|\Phi^{+}_{010}\rangle_S,|\Phi^{-}_{100}\rangle_P|\Phi^{+}_{100}\rangle_S.$ \\

$4$ & $|\Phi^{-}_{000}\rangle_P|\Phi^{-}_{000}\rangle_S,|\Phi^{-}_{000}\rangle_P|\Phi^{-}_{001}\rangle_S,$
      $|\Phi^{-}_{000}\rangle_P|\Phi^{-}_{010}\rangle_S,|\Phi^{-}_{000}\rangle_P|\Phi^{-}_{100}\rangle_S,$ \\&
      $|\Phi^{-}_{001}\rangle_P|\Phi^{-}_{000}\rangle_S,|\Phi^{-}_{001}\rangle_P|\Phi^{-}_{001}\rangle_S,$
      $|\Phi^{-}_{001}\rangle_P|\Phi^{-}_{010}\rangle_S,|\Phi^{-}_{001}\rangle_P|\Phi^{-}_{100}\rangle_S,$ \\ &
      $|\Phi^{-}_{010}\rangle_P|\Phi^{-}_{000}\rangle_S,|\Phi^{-}_{010}\rangle_P|\Phi^{-}_{001}\rangle_S,$
      $|\Phi^{-}_{010}\rangle_P|\Phi^{-}_{010}\rangle_S,|\Phi^{-}_{010}\rangle_P|\Phi^{-}_{100}\rangle_S,$ \\ &
      $|\Phi^{-}_{100}\rangle_P|\Phi^{-}_{000}\rangle_S,|\Phi^{-}_{100}\rangle_P|\Phi^{-}_{001}\rangle_S,$
      $|\Phi^{-}_{100}\rangle_P|\Phi^{-}_{010}\rangle_S,|\Phi^{-}_{100}\rangle_P|\Phi^{-}_{100}\rangle_S.$ \\

\hline
\end{tabular}
\end{table}

\subsection{Complete HGSA for the $N$-photon hyperentanglement}
Our method is also suitable for the complete $N$-photon HGSA, and the setup of our protocol is shown in Fig. 3. Like the complete HBSA and three-photon HGSA, the complete $N$-photon HGSA can be accomplished via three steps.

\begin{figure}
\centering
\includegraphics*[width=0.8\textwidth]{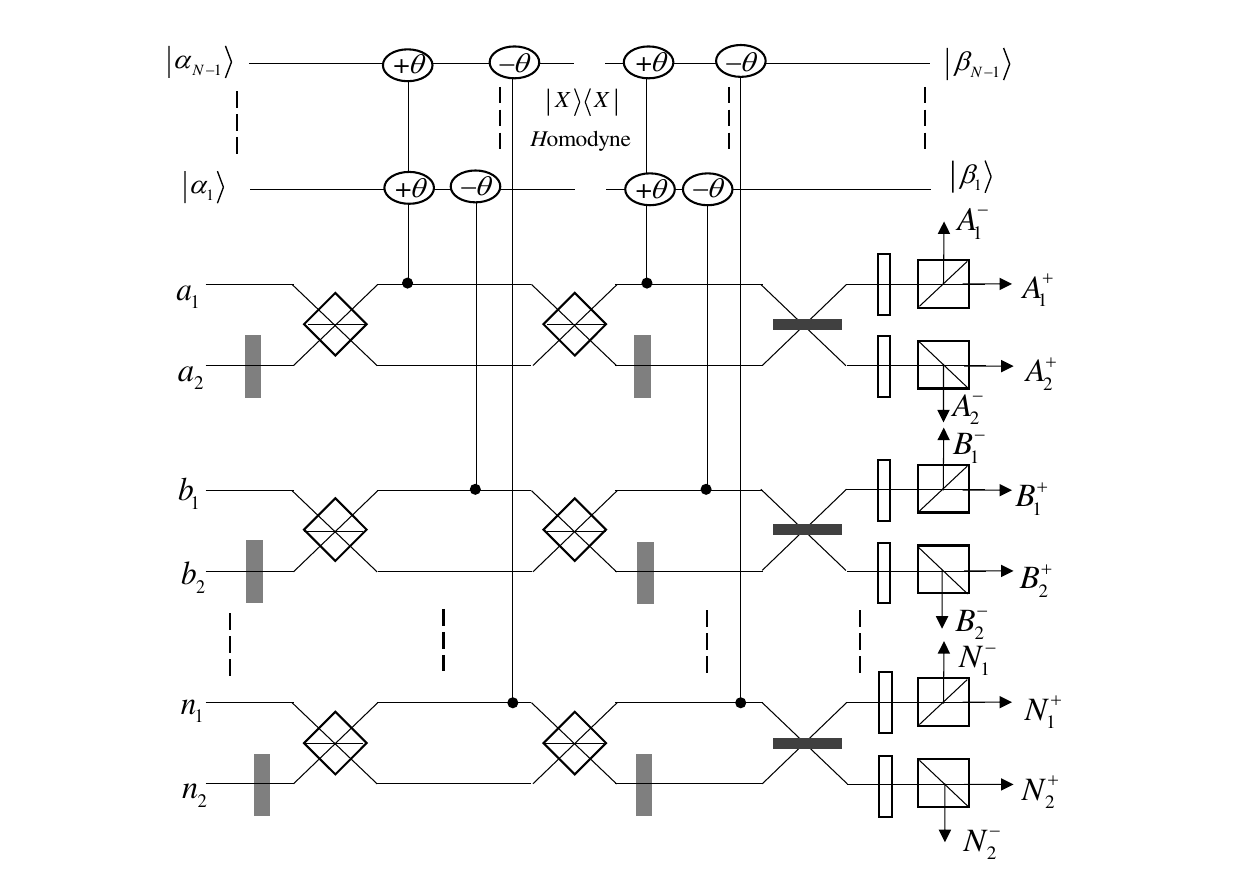}
\caption{Schematic diagram of our complete HGSA scheme for the $N$-photon hyperentanglement. With this device, the $4^N$ hyperentangled GHZ states in two DOFs can be completely distinguished.}
\end{figure}

Firstly, the bit information of polarization entanglement is determined without affecting the spatial-mode entanglement, resorting to the measurements on $N-1$ coherent states. This process can be expressed as
\begin{eqnarray}
&&|\Phi^{\pm}_{ijp\cdots qk}\rangle_{P}|\Phi\rangle_{S}|\alpha_1\rangle|\alpha_2\rangle\cdots|\alpha_{N-2}\rangle|\alpha_{N-1}\rangle  \nonumber \\
&&\rightarrow |\Phi^{\pm}_{ijp\cdots qk}\rangle_{P}|\Phi\rangle_{S}|\alpha_1e^{\pm (i\oplus j)i\theta}\rangle|\alpha_2e^{\pm (i\oplus p)i\theta}\rangle\cdots|\alpha_{N-2}e^{\pm (i\oplus q)i\theta}\rangle|\alpha_{N-1}e^{\pm (i\oplus k)i\theta}\rangle.  \nonumber \\
\end{eqnarray}

Subsequently, the $N$ photons will interact with the other $N-1$ coherent states, and thus the bit information of spatial-mode entanglement is determined. This process can be expressed as
\begin{eqnarray}
&&|\Phi\rangle_{P}|\Phi^{\pm}_{ijp\cdots qk}\rangle_{S}|\beta_1\rangle|\beta_2\rangle\cdots|\beta_{N-2}\rangle|\beta_{N-1}\rangle  \nonumber \\
&&\rightarrow |\Phi\rangle_{P}|\Phi^{\pm}_{ijp\cdots qk}\rangle_{S}|\beta_1e^{\pm (i\oplus j)i\theta}\rangle|\beta_2e^{\pm (i\oplus p)i\theta}\rangle\cdots|\beta_{N-2}e^{\pm (i\oplus q)i\theta}\rangle|\beta_{N-1}e^{\pm (i\oplus k)i\theta}\rangle.  \nonumber \\
\end{eqnarray}
With these two steps, the $4^N$ hyperentangled GHZ states in two DOFs can be separated into $x$ groups,
\begin{eqnarray}
x = 2^{N - 1} * 2^{N - 1} = 4^ {N - 1}.
\end{eqnarray}
Four states exist in each group, and they can be described as $|\Phi^{+}_{ij\cdots k}\rangle_{P}|\Phi^{+}_{ij\cdots k}\rangle_{S}$, $|\Phi^{+}_{ij\cdots k}\rangle_{P}|\Phi^{-}_{ij\cdots k}\rangle_{S}$, $|\Phi^{-}_{ij\cdots k}\rangle_{P}|\Phi^{+}_{ij\cdots k}\rangle_{S}$ and $|\Phi^{-}_{ij\cdots k}\rangle_{P}|\Phi^{-}_{ij\cdots k}\rangle_{S}$. Finally, these four states will be distinguished with linear optics and single photon detectors. Specifically speaking, the number of $|V\rangle$ and number of $|x_2\rangle$ are both even in the detection results for $|\Phi^{+}_{ij\cdots k}\rangle_{P}|\Phi^{+}_{ij\cdots k}\rangle_{S}$, since the phase information of polarization state and spatial-mode state are both "+". For $|\Phi^{+}_{ij\cdots k}\rangle_{P}|\Phi^{-}_{ij\cdots k}\rangle_{S}$, the detection results contain an even number of $|V\rangle$ and an odd number of $|x_2\rangle$. For $|\Phi^{-}_{ij\cdots k}\rangle_{P}|\Phi^{+}_{ij\cdots k}\rangle_{S}$, the detection results contain an odd number of $|V\rangle$ and an even number of $|x_2\rangle$, while the detection results contain an odd number of $|V\rangle$ and an odd number of $|x_2\rangle$ for $|\Phi^{-}_{ij\cdots k}\rangle_{P}|\Phi^{-}_{ij\cdots k}\rangle_{S}$. Up to now, the $4^N$ hyperentangled GHZ states have been completely distinguished.

\section{Discussion and summary}
In the presented schemes, we have proposed an efficient method for the complete two-photon HBSA and multi-photon HGSA with the help of photon number QND and linear optics. The whole discrimination process is accomplished via three steps. Firstly, the bit information of polarization entanglement are obtained by using $N-1$ QNDs for $N$-photon system. Then, other $N-1$ QNDs are required for determining the bit information of spatial-mode entanglement. At last, the relative phase information of hyperentanglement are identified by using the linear optics. With these three steps, both of the complete HBSA and HGSA can be realized. In essence, the quantum manipulation of photonic hyperentangled state in two different DOFs are independent, that's why our schemes can operate successfully.

The cross-Kerr nonlinearity is used for constructing the two-photon QND, which will mainly influence the feasibility of our schemes \cite{Kerr1}. Although it has been widely studied in the past years, the effective nonlinearity strength is still challenging in the single-photon regime with the current technology \cite{Kerr2,Kerr3,Kerr4}. Fortunately, we just need the weak cross-Kerr nonlinearity that could produce the small phase shift in the coherent state. When a sufficiently large amplitude of the coherent state satisfies $\alpha\theta^2\gg1$ ($\theta$ is the cross-phase shift), it is possible to distinguish the small phase of coherent state from the zero phase shift. In our QND detection, the $X$-quadrature measurement is exploited, which leads to an error probability relevant to the strength of coherent state and the phase shift. However, it has shown that the error probability is less than $10^{-5}$ when $\alpha\theta^2 > 9$ \cite{Kerr1}. This shows that our schemes for complete HBSA and HGSA can still operate efficiently in the regime of weak cross-Kerr nonlinearity. Actually, the theoretical and experimental researches have shown that it is promising for us to construct the QND required in our method, resorting to the weak cross-Kerr effect \cite{Kerr5,Kerr6,Kerr7,Kerr8,Kerr9,Kerr10,Kerr11,Kerr12,Kerr13,QND1,QND2,QND3,QND4,QND5}. In addition, some other kinds of interaction can also provide accessible ways to realize the photon number QND, such as cavity-assisted interactions and quantum dot spin in optical microcavity \cite{atom,dot,QD1,QD2,QD3,QD4}.

It is interesting to compare our schemes with the self-assisted schemes, which were proposed by Li \emph{et al.} in 2016 \cite{HBSA2}. In the self-assisted schemes, the spatial-mode entangled states are first distinguished by using the cross-Kerr nonlinearity, completely and nondestructively. Then, the preserved spatial-mode state can be used as the auxiliary entanglement to accomplish the complete BSA or GSA in polarization DOF. However, in our schemes, the bit information of hyperentangled state in both two DOFs are obtained via the QND, and the phase information of hyperentanglement are obtained via the linear optics. Compared with self-assisted mechanism, we just need the achievable small phase shift rather than the giant cross-Kerr effect in the complete HGSA scheme, which makes our method more practical and realizable.

In summary, we have proposed an efficient method for the complete analysis of maximally hyperentangled state in polarization and spatial-mode DOFs via the weak cross-Kerr nonlinearity. This approach is achievable with the current technology, and can be used for the complete HBSA and $N$-photon HGSA. Our method provides an alternative way to the hyperentangled state analysis, and will be useful for the high-capacity quantum communication based on hyperentanglement.


\end{document}